\documentclass[showpacs,amsmath,twocolumn,showkeys,pra]{revtex4-1}
\usepackage{amssymb}
\usepackage{amsbsy}
\usepackage{amsmath}
\usepackage{graphicx}
\usepackage{dcolumn}
\usepackage{bm}
\usepackage{natbib}
\usepackage{bbm}

\newcommand{\ket}[1]{\vert{#1}\rangle}

\usepackage{hyperref}
\usepackage[usenames,dvipsnames]{xcolor}
\definecolor{med-blue}{RGB}{25,25,112}
\hypersetup{colorlinks, linkcolor={red},citecolor={blue}, urlcolor={MidnightBlue}}

\def\be{\begin{equation}}
\def\ee{\end{equation}}

\begin{document}
\title{Engineered Decoherence: Characterization and Suppression}
\author{Swathi S. Hegde and T. S. Mahesh}
\email{mahesh.ts@iiserpune.ac.in}
\affiliation{Department of Physics and NMR Research Center, Indian Institute of Science Education and Research, Pune 411008, India}

\begin{abstract} 
Due to omnipresent environmental interferences, quantum coherences inevitably undergo irreversible transformations over certain time-scales, thus leading to the loss of encoded information. 
This process, known as decoherence, has been a major obstacle in realizing efficient quantum information processors. Understanding the mechanism of decoherence is crucial in developing tools to inhibit it.  Here we utilize a method proposed by Cory and co-workers [Phys. Rev. A {\bf 67}, 062316 (2003)] to engineer artificial decoherence
in the system qubits by randomly perturbing
their surrounding ancilla qubits.  Using a two qubit nuclear magnetic resonance quantum register,
we characterize the artificial decoherence by noise spectroscopy and quantum process tomography.  
Further, we study the efficacy of dynamical decoupling sequences in suppressing the artificial decoherence. 
Here we describe the experimental results and their comparisons with theoretical simulations.
\end{abstract}

\maketitle

\section{Introduction}
The idea of quantum computers or quantum simulators, by harnessing the power of quantum systems, has been seeing a fair advancement towards its practical realization \cite{chuang,shor}. Ideally, in order to accomplish such a hardware, one would prepare a quantum register in a desired initial state, apply a series of quantum operations, and finally read out the output state encoding the information specific to the problem at hand. Neverthless, in reality, a quantum register constantly interacts with its environment, and its evolution may deviate from the ideal path in such a way that a part of its useful information irreversibly leaks into the environment. This process is known as decoherence and is a fundamental threat to quantum computation as well as quantum communication. 
Hence, preserving quantum information against decoherence is an important area of current research.  
Various techniques have already been explored to combat decoherence. These include dynamical decoupling (DD) techniques \cite{cp,uhrig}, post-rectification by quantum error correction \cite{preskill}, use of robust approaches
such as adiabatic quantum computation \cite{farhi}, or encoding quantum information in decoherence-free subspaces \cite{DFS}. Recently DD has received significant attention because of its versatility \cite{lloyd1,lloyd2}. Unlike the other techniques, DD does not require extra qubits, it can be combined with other quantum gates leading to fault tolerant quantum computation \cite{pres,suterprotgates}, and moreover, with the knowledge of noise spectrum, the DD sequences may be further optimized to achieve higher degrees of noise suppression \cite{biercuk1}. 
Above all, engineering decoherence in a controlled way has its own avenues. Wineland and co-workers had earlier demonstrated controlled decoherence of electronic and vibrational degrees in a single trapped ion with the help of random electric fields \cite{myatt}. Cory and co-workers introduced artificial decoherene by using random rf fields on NMR spins \cite{cory}. Such experiments provide insights about decoherence processes and may pave the way in improving decoherence suppression techniques.

A quantum system may dissipate energy, loose coherence, or both during the course of its evolution with the environment. 
Generally, the coherence decays at a rate ($1/T_2$) faster than that of energy dissipation (characterized by $1/T_1$) \cite {biercuk}, thus posing a vital challenge in preserving short lived phase information.
Hence, in this paper, we focus on the phase decoherence, wherein quantum coherence is lost without the loss of energy. We consider a system-environment interaction of the type $\sigma_z\sigma_z$ that leads to the decay of the off-diagonal terms in the system density matrix and thus  damping the phase \cite {zurek}.
In the present work we simulate such a decoherence process on a two qubit nuclear magnetic resonance (NMR) simulator
and describe three aspects of understanding decoherence. They are (i) engineering artificial decoherence,
(ii) characterizing the artificial decoherence by noise spectroscopy (NS) and quantum process tomography (QPT), and 
(iii) suppressing the artificial decoherence using DD.
In the following we briefly introduce to these three aspects.

We follow the methods of engineering artificial decoherence as proposed by Cory and co-workers \cite {cory}. Their decoherence model considers an $N$-dimensional quantum system interacting via $\sigma_z \sigma_z$ interaction with utmost $N^2$-dimensional quantum environment. 
Irreversible phase damping is achieved by constantly perturbing the environment qubits by random classical fields thus mimicking 
a large dimensional environmental bath, and thereby completely 
erasing the information even from the environment qubits. In this work, we utilize a NMR quantum register having two qubits, with one as system and the other as environment.
While introducing the external perturbations on the environment qubit, we characterize the decoherence of the system qubit by NS, i.e., by measuring its noise spectral density \cite{yuge,alvarez}. We also perform QPT of the noise that gives the complete information of the entire noise process for a given duration. Characterizations of decoherence in two different ways have their own advantages: NS helps in extracting the frequency distribution of the noise that may help in further optimizing the DD sequences, and QPT
provides a quantitative estimation of the entire decoherence process.
Finally, we study the performance of two standard DD sequences, CPMG \cite{cp} and Uhrig DD (UDD) \cite{uhrig}, in suppressing artificial decoherence. While perturbations on the environment qubit try to induce decoherence in the system qubit, DD pulses on system qubit attempt to inhibit decoherence. In this sense, it is interesting to study the simultaneous effects of these two competing processes.

The paper proceeds with section II briefly reviewing engineered decoherence with experimental results. 
The pulse-sequences and experimental results of engineered decoherence with CPMG and UDD are discussed in
section III.  Characterization of decoherence with NS and QPT are described in section IV.A and IV.B respectively,
and finally we conclude in section V.

\section{Engineered Decoherence}
\begin{figure}
\centering
\hskip 0.5cm \includegraphics[width=8.7cm]{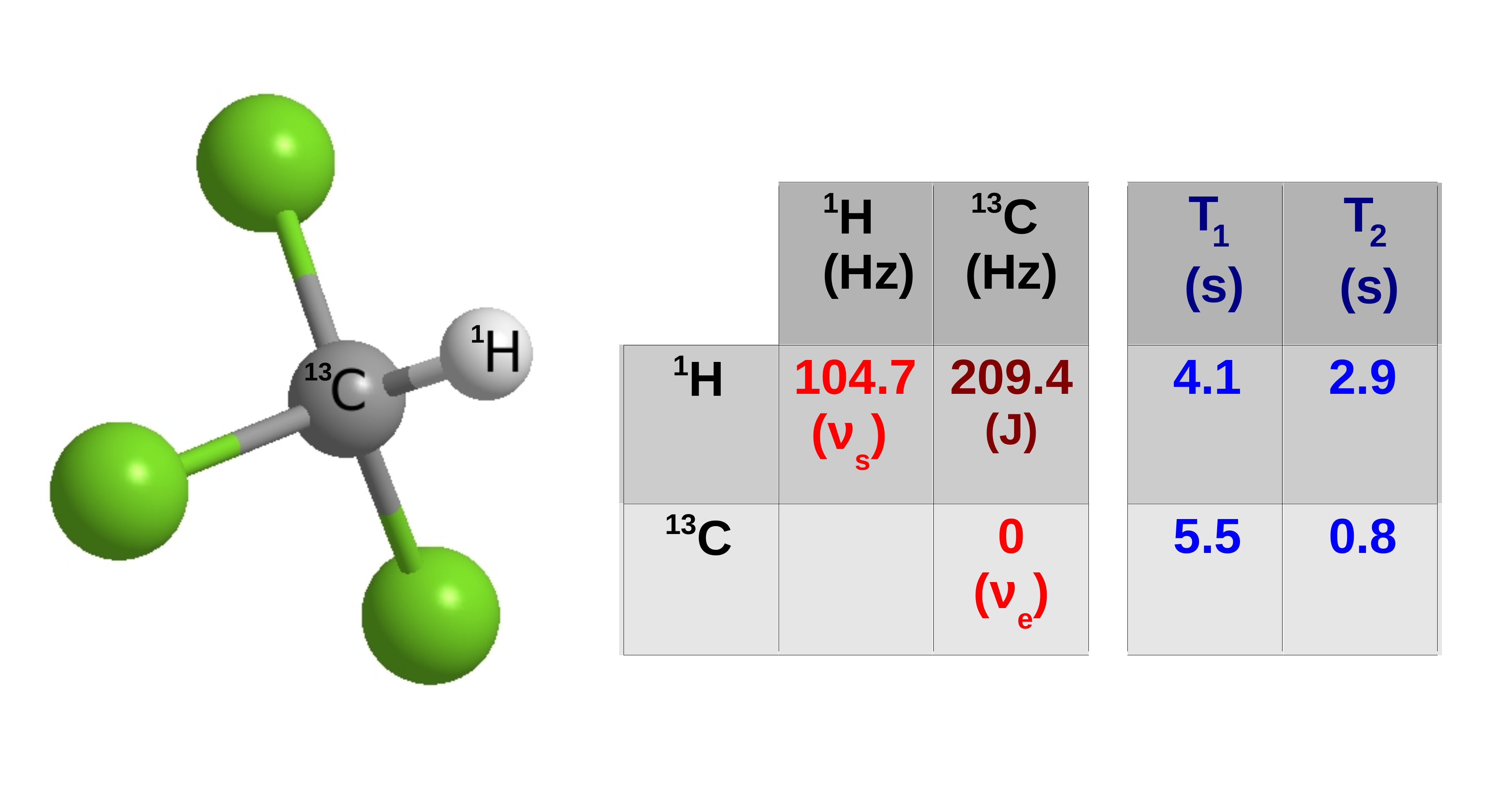}
\caption{
(Color online) The molecular structure of $^{13}$CHCl$_3$ forming a two-qubit NMR quantum register.
The chemical shifts (diagonal elements) and strength of $J$-coupling (off-diagonal element)
are shown in the table above.  The last two columns indicate relaxation time constants $T_1$
and $T_2$ ($T_2$ was measured using a CPMG sequence with $\tau = 3.6$ ms).
}
\label{molecule}
\end{figure}

Here we briefly revise the necessary theory of quantum dynamics under artificial noise as suggested by 
Cory and co-workers \cite{cory}. 
Consider a two qubit system initially in the product state given by,
\begin{equation}
\rho(0) = \rho^s(0) \otimes \rho^e(0),
\end{equation}
where $\rho^s(0)$ is the system state and $\rho^e(0)$ is the environment state. 
In our experiments we have chosen $^1$H and $^{13}$C nuclear spins in $^{13}$C-labelled chloroform ($^{13}$CHCl$_3$ 
dissolved in CDCl$_3$) as the system and environment qubits respectively (Fig. \ref{molecule}).  
Here NMR Hamiltonian is
\begin{equation}
{\cal H}=\pi(\nu_s\sigma_z^{s}+\nu_e\sigma_z^{e}+\frac{J}{2}\sigma_z^{s}\sigma_z^{e}),
\label{Ham}
\end{equation}
where $\nu_s$ and $\nu_e$ are the resonant frequenceis of the system and the environment qubits respectively, $J$ is the coupling strength between the two, and $\sigma_z^{s}$, $\sigma_z^{e}$ are the Pauli operators. 
In a total duration $T$, the propagator $U = e^{-i{\cal H}T}$
entangles the system qubit with the environment qubit via the coupling strength $J$.
Zurek showed that such an interaction leads to the decay of system coherences, with the system populations intact, indicating pure phase damping \cite {zurek}. Further, he proved that an infinite dimensional environment can completely lead to irreversible phase damping, whereas a finite dimensional environment may show a quasiperiodic trend of information exchange between the system and the environment qubits, which may allow the reverse flow of information into the system qubit. Cory and workers cleverly emulated an infinite-dimensional environment by the application of random classical fields on a single environment qubit.  This allows the information to be completely  removed even from the environment qubit, leading to an irreversible flow of information \cite {cory}. 
In NMR, the random classical fields can be realized by a series of kicks, each of which consists of a 
radio-frequency (RF) pulse of an arbitrary angle $\epsilon$. These on-resonant $(\nu_e=0)$ kicks on 
the environment qubit, induce artificial decoherence in the system qubit. The $m^\mathrm{th}$ kick operator is given by 
$K_m = \mathbbm{1}^s\otimes \exp(-i\epsilon_m \sigma_{y}^{e})$,
where $\mathbbm{1}^s$ is the identity operator on system qubit and
the kick-phase is set to $y$ for simplicity. 
Suppose, a total number of $n$ kicks are applied in a duration $T$, with regular intervals $\delta = T/n$, and with a kick rate $\Gamma = n/T$, then the net unitary operator is
\begin{equation}
U_n(T) = K_nU(\delta)K_{n-1}U(\delta) \cdots K_1U(\delta).
\end{equation}
The final combined state is
$\rho(T) = U_n(T)\rho(0)U_n^{\dagger}(T)$,
such that $\rho^s(T) =$ Tr$_e[\rho(T)]$ and $\rho^e(T) =$ Tr$_s[\rho(T)]$. Realization over many random kick-angles $\epsilon$, over an interval $[-\theta,\theta]$, leads to an average behaviour represented by
\begin{equation}
\bar{\rho}^s(T) = \int_{-\theta}^{\theta}\frac{d\epsilon_n}{2\theta}...\int_{-\theta}^{\theta}\frac{d\epsilon_1}{2\theta}\rho^s(T).
\end{equation}
Cory and co-workers proved that \cite{cory}, 
\begin{equation}
\bar{\rho}^s(T) = \sum_{j,k=0,1}f_{jk}(n,T)\rho^s_{jk}(0)|j\rangle\langle k|,
\end{equation}
where $ |j\rangle$, $|k\rangle \in \{\ket{0},\ket{1}\}$  are the eigenstates of $\sigma_z^s$ operator and $f_{jk}(n,T)$ is the decoherence factor given by
\begin{equation}
f_{jk}(n,T) = \mathrm{Tr}_e[{\cal O}^n(\rho^e(0))].
\label{coryt2}
\end{equation} 
Here, the superoperator ${\cal O}$ is neither trace preserving nor Hermitian, and its non-unitary action leading to decoherence is given by
\begin{eqnarray}
{\cal O}^n(\rho^e) &=& c~U_K \rho^e U_K + d ~ \sigma_y U_K \rho^e U_K \sigma_y,
\end{eqnarray}
with $U_K = e^{-i\pi J \delta \sigma_z/2}$, $c+d=1$, $c-d=\gamma$, and $\gamma = \sin(2\theta)/(2\theta)$ \cite{cory}. 
Evidently, the term $f_{01}(n,T)$ containing all the information about dephasing of the system qubit,
depends on the kick-rate $\Gamma = 1/\delta$, range of kick-angles $\theta$, and coupling strength $J$. 
Cory's model predicts that for small kick-angles $\epsilon$ and for lower kick rates $\Gamma$, 
decoherence rate $1/T_2$ increases linearly with $\Gamma$. After a certain value of  $\Gamma$,
$1/T_2$ saturates, and then onwards, it decreases exponentially with $\Gamma$.  
In the latter case, the kicks  actually decouple the environment from the system \cite{cory},
which has similarities with noise decoupling effect \cite{ernst}. 
In our experiments, due to hardware considerations, we focussed only on the low kick-rate regime.

By applying a $(\pi/2)_y$ pulse on the thermal equilibrium state $\mathbbm{1}^s/2+p^s \sigma_z^{s}$, 
we prepared the system qubit in the initial state
$\rho^s(0) = \mathbbm{1}^s/2+p^s \sigma_x^{s}$,
where $p^s \sim 10^{-5}$ is the spin polarization.  
Then we perturbed the environment qubit, initially in the thermal equilibrium state $\rho^e(0) = \mathbbm{1}^e/2+p^e \sigma_z^{e}$, using a series of kicks of random small-angles ($\le 2^\circ$) and random phases (between 0 and $2\pi$).
All the experiments were performed on a Bruker 500 MHz NMR spectrometer at an ambient temperature of 300 K. 

\begin{figure}
\centering
\hspace*{-0.7cm}
\includegraphics[width=10cm]{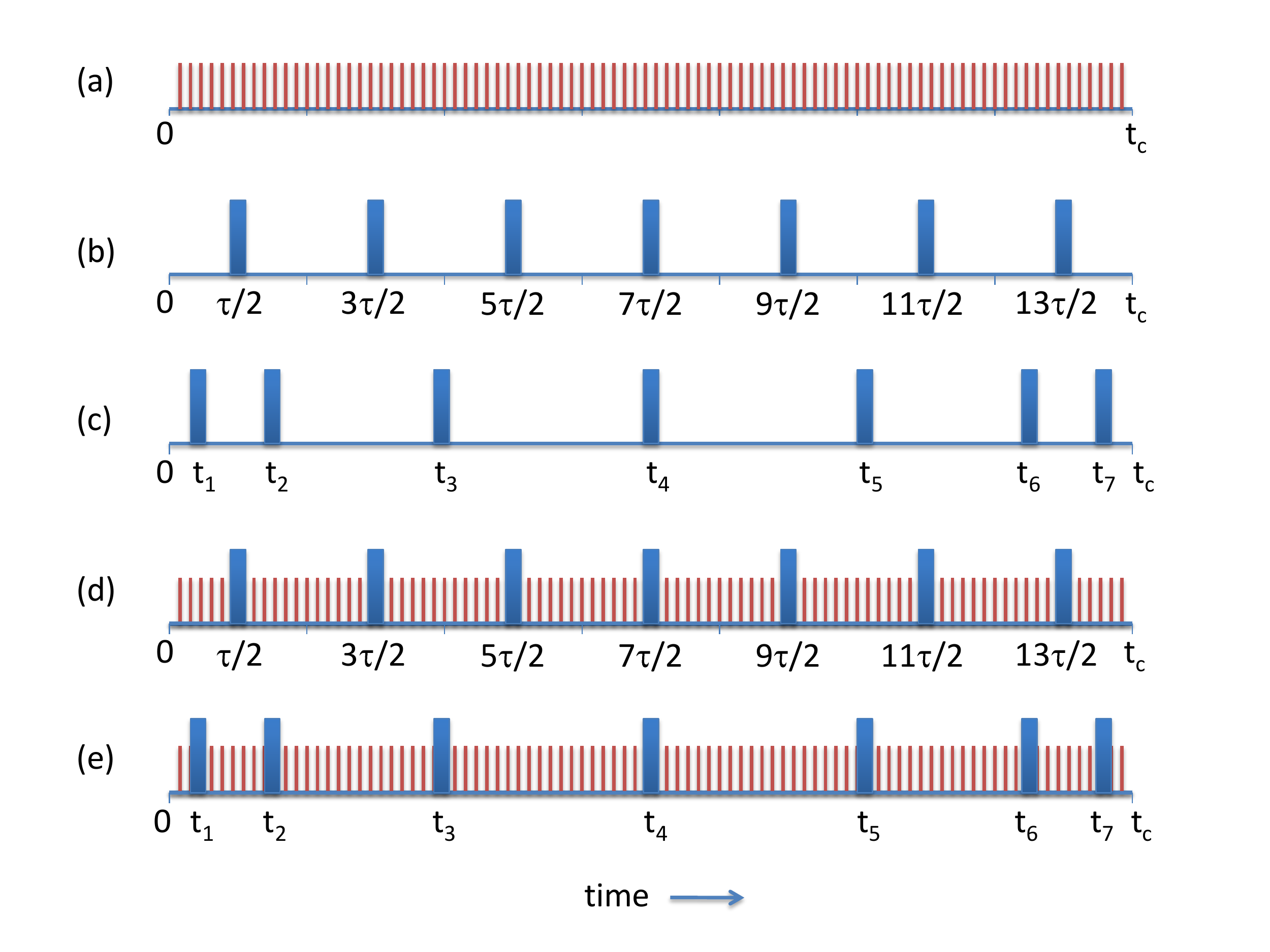}
\caption
{
(Color online) Pulse sequences for (a) only kicks, (b) CPMG, (c) 7-pulse UDD, (d) kicks with CPMG, and (e) kicks with UDD.
The kicks applied on the environment qubit ($^{13}$C) are shown by thin, short vertical lines,  and
the DD $\pi$ pulses on the system qubit ($^1$H) are shown by filled rectangles.  
The time instants of $\pi$ pulses for the first cycle of duration $t_c = 7 \tau$ are shown.  
In the case of UDD (c and e), exact time instants are calculated from eqn. \ref{uddeq}.
}
\label{pulseq}
\end{figure}

\begin{figure}
\centering
\hspace*{-.6cm}
\includegraphics[width=8cm]{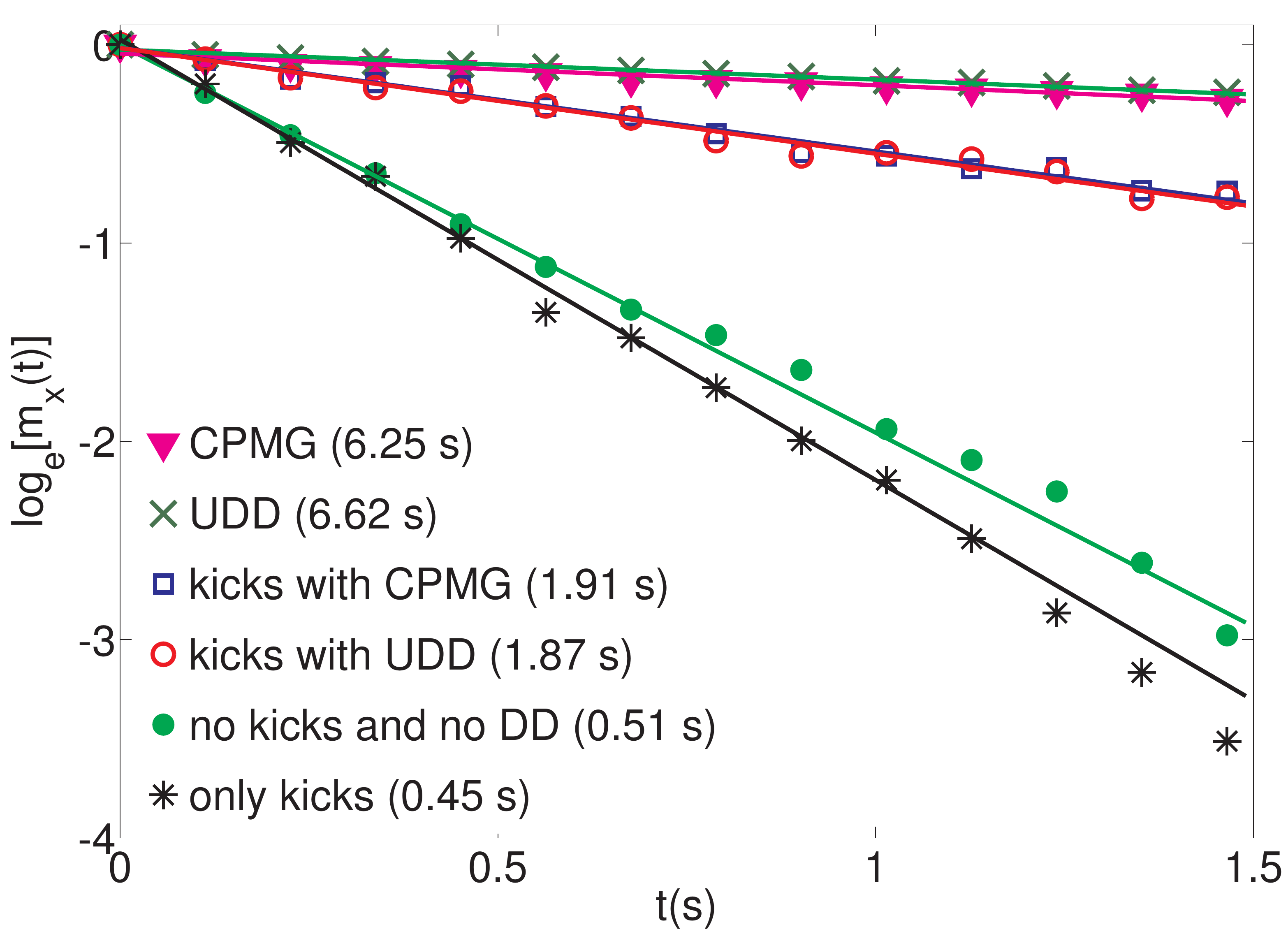}
\caption{
(Color online) Logarithm of transverse magnetization $\log(M_x)$ as a function of time under different 
pulse sequences as indicated in the legend. Here $\tau = 3.2$ ms,
$\Gamma = 25$ kicks/ms, and $\epsilon \in [0^\circ,1^\circ]$. The $T_2$ values for various pulse sequences are shown in the legend.
}
\label{comp_m}
\end{figure}

The experimental scheme for realizing engineered decoherence is shown in Fig. \ref{pulseq}(a).
The decay of the system coherence is monitored by measuring its transverse magnetization 
$M_x(m t_c) = Tr[\rho^s(m t_c) \sigma_x^s] $ after $m$ cycles each of duration $t_c$.
The results of the experiment for $\epsilon \in [0^\circ,1^\circ]$ and $\Gamma = 25$ kicks/ms
are shown in Fig. \ref{comp_m} (indicated by stars).  
For comparison we have also shown the decay of magnetization without kicks (indicated by filled circles).  
Evidently, the kicks on environment have introduced additional decoherence, apart from the natural relaxation processes, thereby increasing 
the decay of system coherence.

In the following we briefly describe two standard DD sequences and then proceed to the characterization
of engineered decoherence.

\section{Dynamical Decoupling Techniques}
Dynamical decoupling attempts to inhibit decoherence of system by 
rapid modulation of the system state so as to cancel the system-environment
joint evolutions.  In this section, we briefly review two standard 
DD sequences: (i) CPMG \cite{cp} and (ii) UDD \cite{uhrig}.
CPMG consists of a series of equidistant $\pi$ pulses applied on the system qubits (Fig. \ref{pulseq}(b)).
In general, smaller the duration $\tau$ between the $\pi$ pulses, larger
the bandwidth of noise that is suppressed, and thereby increasing the efficiency of DD.  
In practice, the phases are chosen such that the initial state is stationary under the pulses,
so that the DD sequence is robust against pulse errors.
As described before, in our experiments, the system qubit is initialized to $\sigma_x^s$ 
(traceless-deviation) and all the $\pi$ pulses are applied about $x$-axis.

While CPMG is very efficient in suppressing
decoherence from a noise-spectrum with soft frequency cut-off, Uhrig found
an optimum sequence, \textit{viz}. UDD, for sharp frequency cut-off \cite{uhrig}.  
In an N-pulse UDD of cycle time $t_c$,  the time instant $t_j$ of $j^\mathrm{th}$ $\pi$-pulse is given by 
(Fig. \ref{pulseq}(c))
\begin{equation}
t_j = t_c \sin^2 \left[\frac{\pi j}{2(N+1)}\right].
\label{uddeq}
\end{equation}

Experiments were performed with ony DD (Fig. \ref{pulseq}(b), (c)) and kicks with DD (Fig. \ref{pulseq}(d), (e)). In the latter case, the kicks were applied on the environment qubits and the simultaneous DD pulses were applied on the system qubits. The results of the experiments for $t_c = 22.4$ ms and with different
kick-parameters are shown in Fig. \ref{comp_m}.  The competition between
kicks-induced decoherence and DD sequences can be readily observed (indicated by squares and open circles).
In this case, the performance of CPMG and UDD sequences are almost identical.
It may be noted that detailed comparative studies of CPMG and UDD in preserving various
initial states under natural relaxation processes have been studied elsewhere 
\cite{ashok,soumyaudd}.

\section{Characterization of Engineered Decoherence}
\subsection{Noise Spectroscopy}
Noise Spectroscopy (NS) provides information about noise spectral density, which is the frequency distribution of noise and is helpful in not only understanding the performance of standard DD sequences, but also in optimizing them \cite{biercuk,biercuk1,pan}. 
It was observed independently by Yuge et al \cite{yuge} and Alvarez et al \cite{alvarez} that since
the efficiency of CPMG depends on the rate ($1/\tau$) of $\pi$ pulses, the information of the noise spectral density is contained in the decoherence rate $1/T_2(\omega)$, where $\omega = \pi/\tau$. 
A given CPMG sequence, with a particular $\tau$ value, has a  sinc-like filter-function \cite{biercuk}. 
\begin{figure}[t]
\centering
\includegraphics[width=8.7cm]{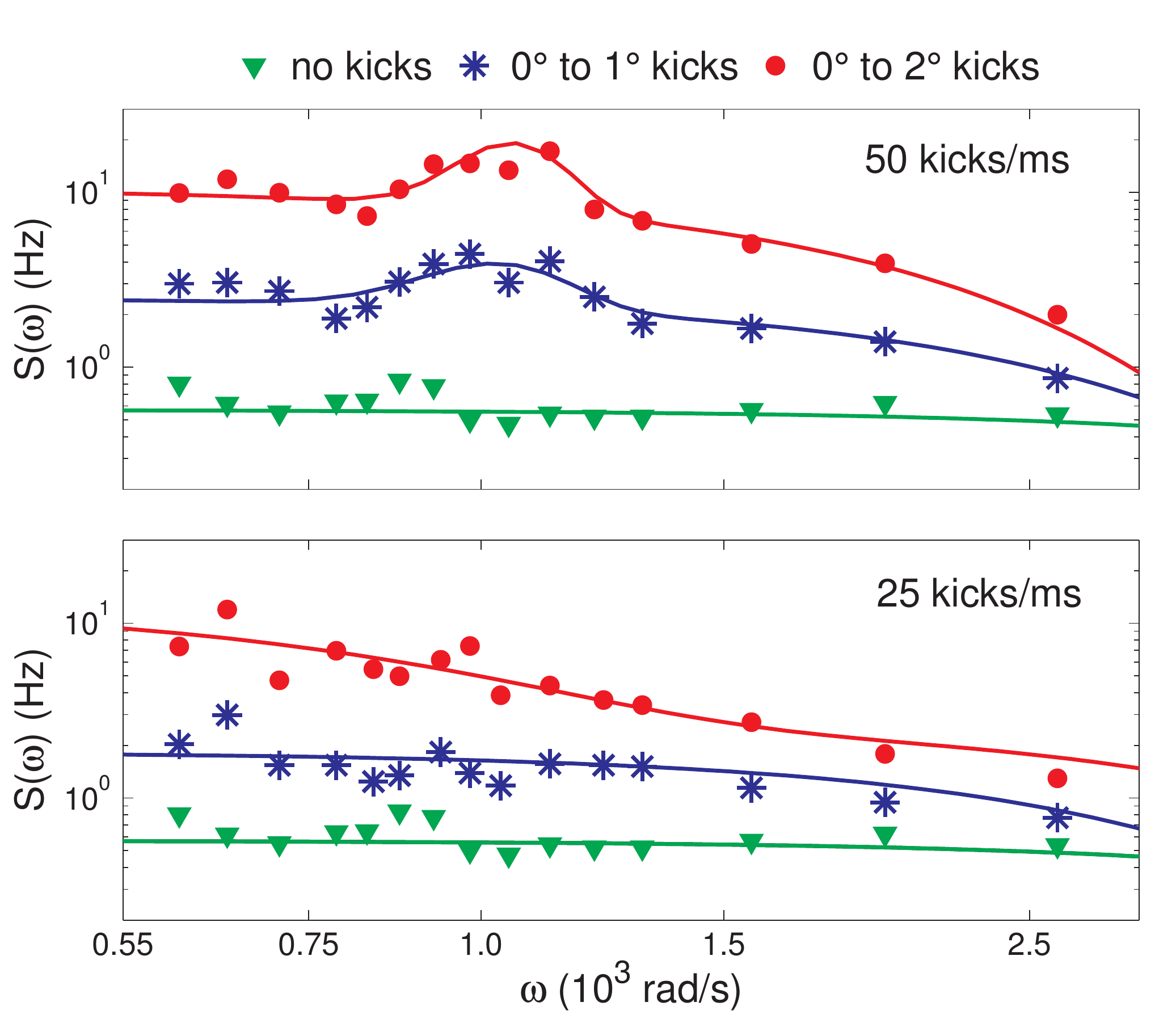}
\caption{
(Color online) Experimental spectral density profiles with different kick-angles
(as indicated in the legend) and with kick-rates 50 kicks/ms (top trace)
and 25 kicks/ms (bottom trace). In both the traces, experimental spectral
density profile without kicks is also shown for comparison. The smooth
lines correspond to fits with one or two Gaussians.
}
\label{sd}
\end{figure}
In the limit of a large number of $\pi$ pulses, the filter function resembles a delta peak at $\omega$, and samples this particular spectral frequency. The amplitude of the noise $S(\omega)$ can be determined by 
using the relation \cite{yuge}
\begin{eqnarray}
S(\omega) \simeq \frac{\pi^2}{4 T_2(\omega)}.
\label{sdt2}
\end{eqnarray}
Thus by measuring $T_2(\omega)$ for a range of $\omega=\pi/\tau$ values, we can scan the
profile of $S(\omega)$. 

The pulse sequences for measuring noise spectral density without and with kicks are
as shown in Figs. \ref{pulseq}(b) and (d) respectively.  
We measured the decay constant $T_2(\omega)$ of the system qubit by measuring
transverse magnetization after time intervals $mt_c$ and fitting the decaying signal to
exponential. 

The experimental spectral density profiles, obtained using eqn. \ref{sdt2},
of only natural decoherence (lowest curve in each sub-plot), and with kicks of
different kick-parameters are shown in Fig. \ref{sd}.
Clearly the effect of kicks is to increase the area under the spectral
density profiles and thereby leading to faster decoherence.  Moreover, 
for a given kick-rate $\Gamma$, larger the range of kick-angles, higher is the 
spectral density profile.  Interestingly, we observe some characteristic
features in the noise spectral density at higher kick-rate (50 kicks/ms).  Similar
features were earlier observed by Suter and co-workers due to a decoupling
sequence being applied on environment spins \cite{alvarez}.
It was also predicted by Cory and co-workers that at very
high $\Gamma$, the kicks actually tend to decouple the system from the environment
\cite{cory}.

\begin{figure}
\centering
\includegraphics[width=6cm]{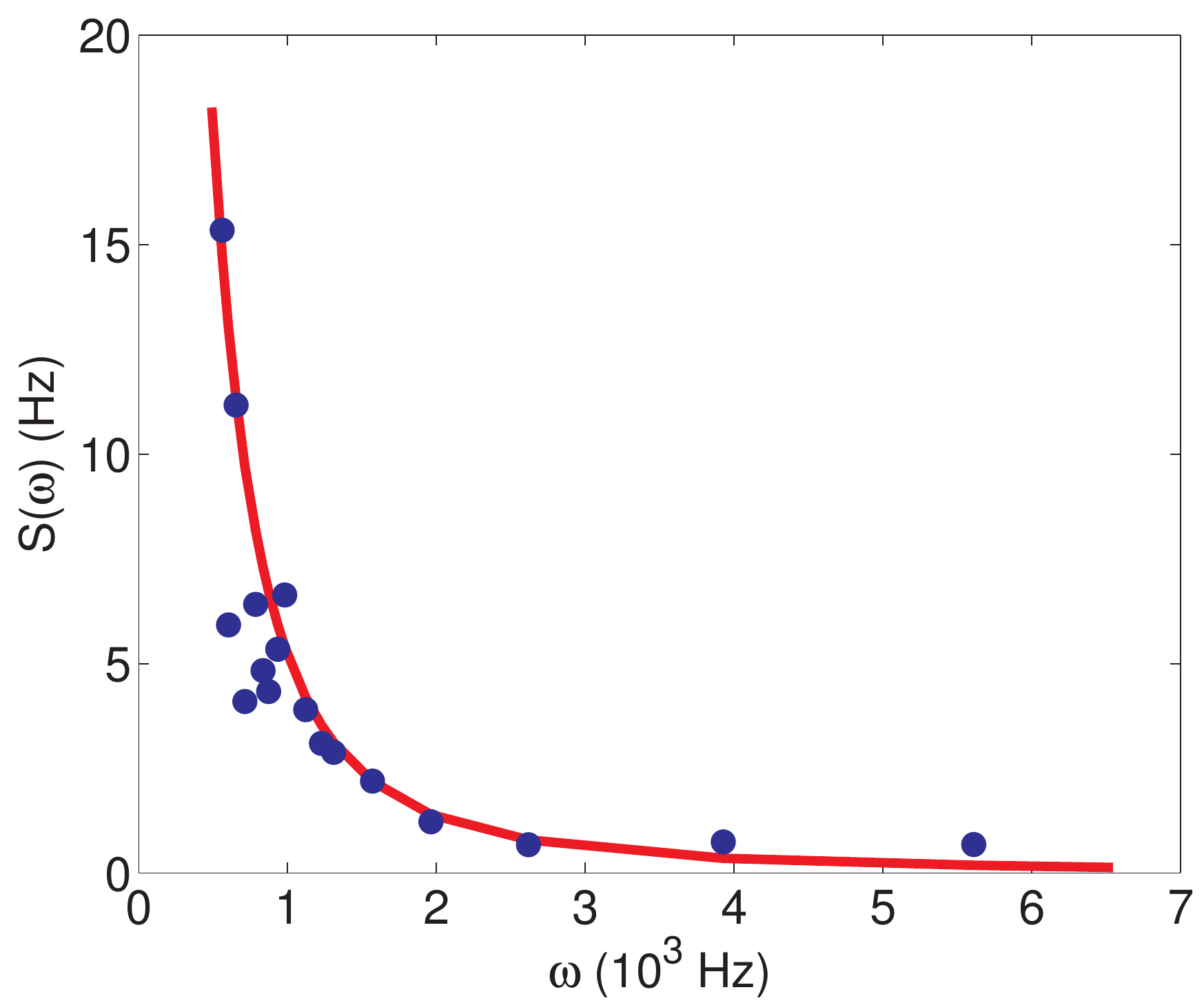}
\caption{
(Color online) The experimental spectral density (dots) with kick-rate $\Gamma = 25$ kicks/ms and 
kick-angles $\epsilon \in [0^\circ,2^\circ]$. Theoretical spectral density 
simulated from Cory's model for the same kick-parameters is shown by smooth line.
}
\label{sd_Th}
\end{figure}

One can also extract the kicks-contributed spectral density 
from Cory's model by calculating the theoretical $T_2(\omega)$ (using Eqns. \ref{coryt2} and \ref{sdt2}) 
at various $\omega$ values, and then compare with
the experimental spectral density for the corresponding kick-parameters.
This comparison is shown in Fig. \ref{sd_Th} for kick-rate of $25$ kicks/ms
and kick-angles in the range 0 to 2 degrees.  
To obtain the experimental kicks-contributed spectral density alone, we have
subtracted the intrinsic spectral density of the system qubit (with no kicks) 
from the total spectral density with kicks. 
The reasonable agreement between the simulated curve and the experimental data
confirms the relevance of Cory's decoherence model in this regime.

\subsection{Quantum Process Tomography}
Another way to characterize decoherence is by using quantum process tomography (QPT),
which determines the entire process acting on a system \cite{chuang}.
Consider a process $\mathcal{E}$ acting on an initial system state $\rho$ that transforms it to a final state 
$\rho'  = \mathcal{E}(\rho)$.  Expressing in Kraus basis $\{{\tilde E}_m\}$, we get
\begin{eqnarray}
\mathcal{E}(\rho) = \sum_{mn} {\tilde E}_m \rho {\tilde E}_n^{\dagger} \chi_{mn}, 
\end{eqnarray}
wherein the matrix elements $\chi_{mn}$ completely characterize the process. In our case, the $\chi$ matrix corresponds to the kick-induced noise process.
Single qubit QPT involves applying the process
$\mathcal{E}$ independently on four initial states $\{\rho_j\}$ that form a complete set,
measuring the corresponding final states $\{\rho_j'\}$ by quantum state tomography,
and evaluating the $\chi$ matrix using the relations \cite{chuang}
\begin{eqnarray}
\lambda{jk} &=& Tr[\rho_j'\rho_k], \nonumber \\
\beta_{jk}^{pq} &=& Tr[\tilde{E}_p\rho_j\tilde{E}^\dagger_q\rho_k], ~ \mathrm{and} \nonumber \\
\sum_{pq} \beta_{jk}^{pq} \chi_{pq} &=& \lambda_{jk}.
\end{eqnarray}

\begin{figure}[t]
\centering
\hspace*{-0.5cm}
\includegraphics[width=9cm]{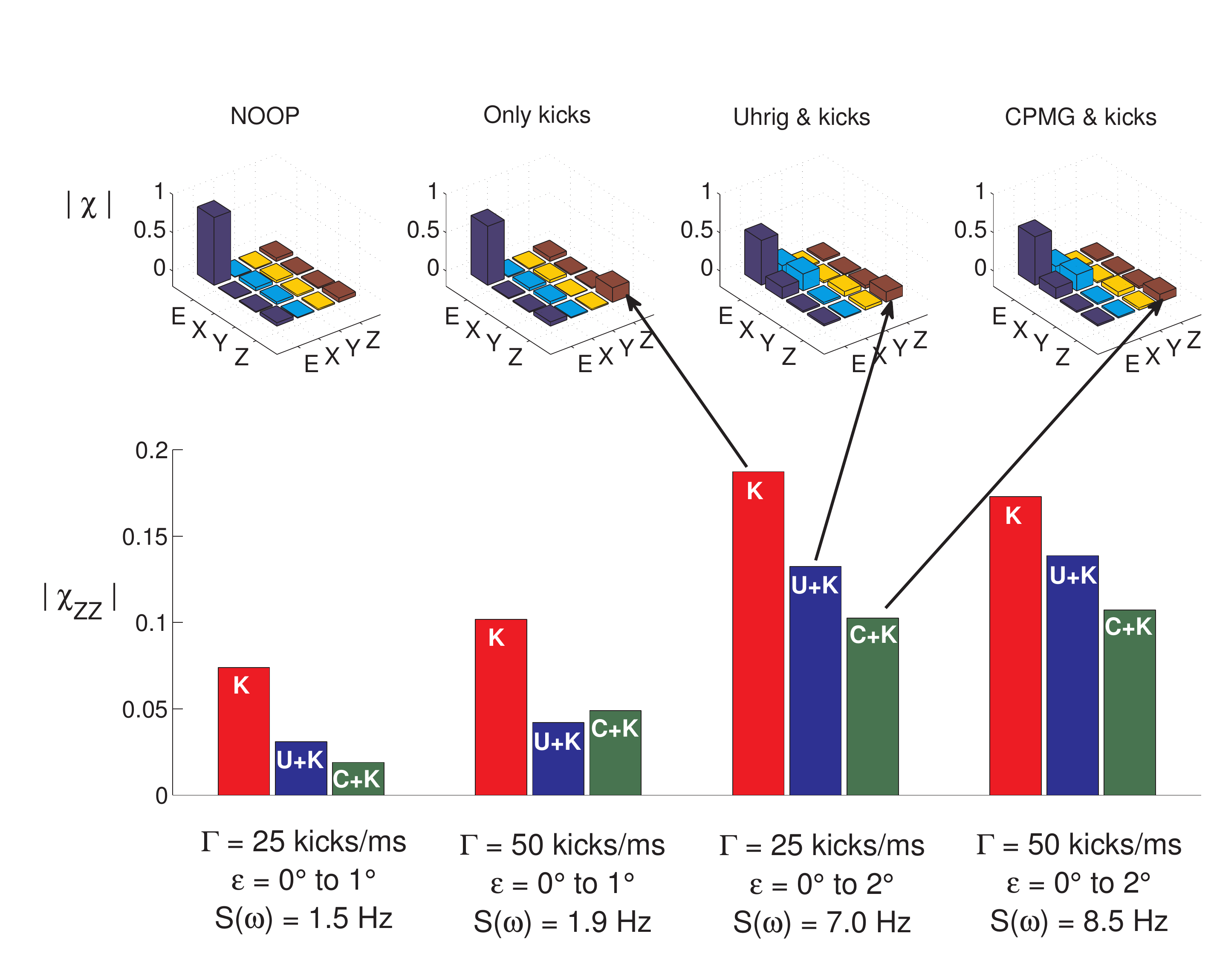}
\caption{
(Color online) The $ZZ$-components of experimental $\vert \chi \vert$ matrices obtained
under kicks (K), UDD with kicks (U+K), and CPMG with kicks (C+K) are shown in the lower bar plots.
In both CPMG and UDD, the cycle time is set to $t_c = 7\tau = 28$ ms.
Corresponding experimental spectral densities $S(\omega)$ 
(where $\omega = \pi/\tau = 785 ~\mathrm{rad/s})$ 
and other kick-parameters are as indicated in the figure.
Full $\vert \chi \vert$ matrices for a particular case (as indicated by the arrows) along with the NOOP case, 
are shown in the upper trace.
}
\label{qpt}
\end{figure}

We measured $\chi$ matrices with various kick-parameters and DD sequences as summarized in Fig. \ref{qpt}.
The absolute of $\chi$ matrices for identity process (NOOP), only kicks, Uhrig with kicks, and CPMG with kicks 
(for $\Gamma = 25$ kicks/ms, $\epsilon \in [0^\circ,2^\circ]$, $\tau = 4$ ms) are shown in the upper trace of Fig. \ref{qpt}.
The identity process is carried out with a Hahn-echo type sequence to refocus the internal Hamiltonian.
The $\chi_{ZZ} $ element, which encodes the effect of dephasing noise, is absent in NOOP, maximum when only kicks are applied, and partially suppressed under the DD sequences. While the DD sequences partially suppress the decoherence,
the non-idealities in the $\pi$-pulses introduce NOT operations (XX, EX, and XE-processes) as seen in the bar plots.
The $\vert \chi_{ZZ} \vert$ elements under different kick-parameters and DD sequences with $\tau = 4$ ms ($\omega = \pi/\tau = 785$ rad/s) are compared in the lower trace of Fig. \ref{qpt}.  The spectral densities at 785 rad/s are indicated in each case.
Clearly, kicks with $\epsilon \in [0^\circ,2^\circ]$ produce much stronger decoherence than
those with $\epsilon \in [0^\circ,1^\circ]$, as can be observed by relative values of spectral densities as well as 
those of $\vert \chi_{ZZ} \vert$ values.  While both CPMG and UDD suppress the decoherence to some extent, 
CPMG seems to be having an overall higher efficiency than UDD, as expected from the broad Gaussian spectral density profiles.

\section{Conclusions}
Quantum devices, albeit their novel applications, suffer from
intrinsic decoherence caused by environmental noise.  
Engineering artificial quantum noise may play an important role in developing
tools to suppress decoherence.
In this work we used the method proposed by 
Cory and co-workers \cite{cory} to engineer decoherence on a two-qubit
NMR quantum register.  The method involves applying a series of random
small-angle kicks on the environment qubit to induce decoherence in the
system qubit.  We characterized the kick-induced decoherence by noise
spectroscopy and quantum process tomography.  While the noise spectroscopy
provided the detailed spectral density profiles, the quantum process tomography
revealed the overall phase decoherence acted on the system qubit.  
These characterizations provided a better understanding of kick-induced decoherence.
We also studied the efficiency of standard dynamical decoupling methods,
viz., CPMG and UDD, in suppressing the engineered decoherence.  CPMG 
seemed to have an over-all better performance as expected from the broad
spectral density profiles revealed by noise spectroscopy.
We believe that studying the simultaneous effects of kicks inducing decoherence
and dynamical decoupling suppressing decoherence may provide an important avenue
for designing new robust optimized dynamical decoupling sequences.

\section*{Acknowledgements}
The authors are grateful to Mr. Abhishek Shukla and 
Prof. Anil Kumar for discussions.  This work was partly supported by 
DST project SR/S2/LOP-0017/2009.

\bibliographystyle{apsrev4-1} 
\bibliography{eng_dec1}

\begin{thebibliography}{22}%
\makeatletter
\providecommand \@ifxundefined [1]{%
 \@ifx{#1\undefined}
}%
\providecommand \@ifnum [1]{%
 \ifnum #1\expandafter \@firstoftwo
 \else \expandafter \@secondoftwo
 \fi
}%
\providecommand \@ifx [1]{%
 \ifx #1\expandafter \@firstoftwo
 \else \expandafter \@secondoftwo
 \fi
}%
\providecommand \natexlab [1]{#1}%
\providecommand \enquote  [1]{``#1''}%
\providecommand \bibnamefont  [1]{#1}%
\providecommand \bibfnamefont [1]{#1}%
\providecommand \citenamefont [1]{#1}%
\providecommand \href@noop [0]{\@secondoftwo}%
\providecommand \href [0]{\begingroup \@sanitize@url \@href}%
\providecommand \@href[1]{\@@startlink{#1}\@@href}%
\providecommand \@@href[1]{\endgroup#1\@@endlink}%
\providecommand \@sanitize@url [0]{\catcode `\\12\catcode `\$12\catcode
  `\&12\catcode `\#12\catcode `\^12\catcode `\_12\catcode `\%12\relax}%
\providecommand \@@startlink[1]{}%
\providecommand \@@endlink[0]{}%
\providecommand \url  [0]{\begingroup\@sanitize@url \@url }%
\providecommand \@url [1]{\endgroup\@href {#1}{\urlprefix }}%
\providecommand \urlprefix  [0]{URL }%
\providecommand \Eprint [0]{\href }%
\providecommand \doibase [0]{http://dx.doi.org/}%
\providecommand \selectlanguage [0]{\@gobble}%
\providecommand \bibinfo  [0]{\@secondoftwo}%
\providecommand \bibfield  [0]{\@secondoftwo}%
\providecommand \translation [1]{[#1]}%
\providecommand \BibitemOpen [0]{}%
\providecommand \bibitemStop [0]{}%
\providecommand \bibitemNoStop [0]{.\EOS\space}%
\providecommand \EOS [0]{\spacefactor3000\relax}%
\providecommand \BibitemShut  [1]{\csname bibitem#1\endcsname}%
\let\auto@bib@innerbib\@empty
\bibitem [{\citenamefont {Nielsen}\ and\ \citenamefont
  {Chuang}(2010)}]{chuang}%
  \BibitemOpen
  \bibfield  {author} {\bibinfo {author} {\bibfnamefont {M.~A.}\ \bibnamefont
  {Nielsen}}\ and\ \bibinfo {author} {\bibfnamefont {I.~L.}\ \bibnamefont
  {Chuang}},\ }\href@noop {} {\emph {\bibinfo {title} {Quantum computation and
  quantum information}}}\ (\bibinfo  {publisher} {Cambridge university press},\
  \bibinfo {year} {2010})\BibitemShut {NoStop}%
\bibitem [{\citenamefont {Shor}(1997)}]{shor}%
  \BibitemOpen
  \bibfield  {author} {\bibinfo {author} {\bibfnamefont {P.~W.}\ \bibnamefont
  {Shor}},\ }\href@noop {} {\bibfield  {journal} {\bibinfo  {journal} {SIAM J.
  Sci. Statist. Comput.}\ }\textbf {\bibinfo {volume} {26}},\ \bibinfo {pages}
  {1484} (\bibinfo {year} {1997})}\BibitemShut {NoStop}%
\bibitem [{\citenamefont {Meiboom}\ and\ \citenamefont {Gill}(1958)}]{cp}%
  \BibitemOpen
  \bibfield  {author} {\bibinfo {author} {\bibfnamefont {S.}~\bibnamefont
  {Meiboom}}\ and\ \bibinfo {author} {\bibfnamefont {D.}~\bibnamefont {Gill}},\
  }\href {\doibase http://dx.doi.org/10.1063/1.1716296} {\bibfield  {journal}
  {\bibinfo  {journal} {Rev. of Sci. Instrum.}\ }\textbf {\bibinfo {volume}
  {29}},\ \bibinfo {pages} {688} (\bibinfo {year} {1958})}\BibitemShut
  {NoStop}%
\bibitem [{\citenamefont {Uhrig}(2007)}]{uhrig}%
  \BibitemOpen
  \bibfield  {author} {\bibinfo {author} {\bibfnamefont {G.~S.}\ \bibnamefont
  {Uhrig}},\ }\href {\doibase 10.1103/PhysRevLett.98.100504} {\bibfield
  {journal} {\bibinfo  {journal} {Phys. Rev. Lett.}\ }\textbf {\bibinfo
  {volume} {98}},\ \bibinfo {pages} {100504} (\bibinfo {year}
  {2007})}\BibitemShut {NoStop}%
\bibitem [{\citenamefont {Preskill}(1998)}]{preskill}%
  \BibitemOpen
  \bibfield  {author} {\bibinfo {author} {\bibfnamefont {J.}~\bibnamefont
  {Preskill}},\ }\href@noop {} {\bibfield  {journal} {\bibinfo  {journal}
  {Proc. R. Soc. Lond. A}\ }\textbf {\bibinfo {volume} {454}},\ \bibinfo
  {pages} {385} (\bibinfo {year} {1998})}\BibitemShut {NoStop}%
\bibitem [{\citenamefont {Farhi}\ \emph {et~al.}(2000)\citenamefont {Farhi},
  \citenamefont {Goldstone}, \citenamefont {Gutmann},\ and\ \citenamefont
  {Sipser}}]{farhi}%
  \BibitemOpen
  \bibfield  {author} {\bibinfo {author} {\bibfnamefont {E.}~\bibnamefont
  {Farhi}}, \bibinfo {author} {\bibfnamefont {J.}~\bibnamefont {Goldstone}},
  \bibinfo {author} {\bibfnamefont {S.}~\bibnamefont {Gutmann}}, \ and\
  \bibinfo {author} {\bibfnamefont {M.}~\bibnamefont {Sipser}},\ }\href@noop {}
  {\bibfield  {journal} {\bibinfo  {journal} {arXiv:quant-ph/0001106}\ }
  (\bibinfo {year} {2000})}\BibitemShut {NoStop}%
\bibitem [{\citenamefont {Lidar}\ and\ \citenamefont {Whaley}(2003)}]{DFS}%
  \BibitemOpen
  \bibfield  {author} {\bibinfo {author} {\bibfnamefont {D.~A.}\ \bibnamefont
  {Lidar}}\ and\ \bibinfo {author} {\bibfnamefont {K.~B.}\ \bibnamefont
  {Whaley}},\ }in\ \href@noop {} {\emph {\bibinfo {booktitle} {Irreversible
  Quantum Dynamics}}}\ (\bibinfo  {publisher} {Springer},\ \bibinfo {year}
  {2003})\ pp.\ \bibinfo {pages} {83--120}\BibitemShut {NoStop}%
\bibitem [{\citenamefont {Viola}\ \emph {et~al.}(1999)\citenamefont {Viola},
  \citenamefont {Knill},\ and\ \citenamefont {Lloyd}}]{lloyd1}%
  \BibitemOpen
  \bibfield  {author} {\bibinfo {author} {\bibfnamefont {L.}~\bibnamefont
  {Viola}}, \bibinfo {author} {\bibfnamefont {E.}~\bibnamefont {Knill}}, \ and\
  \bibinfo {author} {\bibfnamefont {S.}~\bibnamefont {Lloyd}},\ }\href
  {\doibase 10.1103/PhysRevLett.82.2417} {\bibfield  {journal} {\bibinfo
  {journal} {Phys. Rev. Lett.}\ }\textbf {\bibinfo {volume} {82}},\ \bibinfo
  {pages} {2417} (\bibinfo {year} {1999})}\BibitemShut {NoStop}%
\bibitem [{\citenamefont {Viola}\ and\ \citenamefont {Lloyd}(1998)}]{lloyd2}%
  \BibitemOpen
  \bibfield  {author} {\bibinfo {author} {\bibfnamefont {L.}~\bibnamefont
  {Viola}}\ and\ \bibinfo {author} {\bibfnamefont {S.}~\bibnamefont {Lloyd}},\
  }\href {\doibase 10.1103/PhysRevA.58.2733} {\bibfield  {journal} {\bibinfo
  {journal} {Phys. Rev. A}\ }\textbf {\bibinfo {volume} {58}},\ \bibinfo
  {pages} {2733} (\bibinfo {year} {1998})}\BibitemShut {NoStop}%
\bibitem [{\citenamefont {Ng}\ \emph {et~al.}(2011)\citenamefont {Ng},
  \citenamefont {Lidar},\ and\ \citenamefont {Preskill}}]{pres}%
  \BibitemOpen
  \bibfield  {author} {\bibinfo {author} {\bibfnamefont {H.~K.}\ \bibnamefont
  {Ng}}, \bibinfo {author} {\bibfnamefont {D.~A.}\ \bibnamefont {Lidar}}, \
  and\ \bibinfo {author} {\bibfnamefont {J.}~\bibnamefont {Preskill}},\ }\href
  {\doibase 10.1103/PhysRevA.84.012305} {\bibfield  {journal} {\bibinfo
  {journal} {Phys. Rev. A}\ }\textbf {\bibinfo {volume} {84}},\ \bibinfo
  {pages} {012305} (\bibinfo {year} {2011})}\BibitemShut {NoStop}%
\bibitem [{\citenamefont {Zhang}\ \emph {et~al.}(2014)\citenamefont {Zhang},
  \citenamefont {Souza}, \citenamefont {Brandao},\ and\ \citenamefont
  {Suter}}]{suterprotgates}%
  \BibitemOpen
  \bibfield  {author} {\bibinfo {author} {\bibfnamefont {J.}~\bibnamefont
  {Zhang}}, \bibinfo {author} {\bibfnamefont {A.~M.}\ \bibnamefont {Souza}},
  \bibinfo {author} {\bibfnamefont {F.~D.}\ \bibnamefont {Brandao}}, \ and\
  \bibinfo {author} {\bibfnamefont {D.}~\bibnamefont {Suter}},\ }\href
  {\doibase 10.1103/PhysRevLett.112.050502} {\bibfield  {journal} {\bibinfo
  {journal} {Phys. Rev. Lett.}\ }\textbf {\bibinfo {volume} {112}},\ \bibinfo
  {pages} {050502} (\bibinfo {year} {2014})}\BibitemShut {NoStop}%
\bibitem [{\citenamefont {Biercuk}\ \emph {et~al.}(2009)\citenamefont
  {Biercuk}, \citenamefont {Uys}, \citenamefont {VanDevender}, \citenamefont
  {Shiga}, \citenamefont {Itano},\ and\ \citenamefont {Bollinger}}]{biercuk1}%
  \BibitemOpen
  \bibfield  {author} {\bibinfo {author} {\bibfnamefont {M.~J.}\ \bibnamefont
  {Biercuk}}, \bibinfo {author} {\bibfnamefont {H.}~\bibnamefont {Uys}},
  \bibinfo {author} {\bibfnamefont {A.~P.}\ \bibnamefont {VanDevender}},
  \bibinfo {author} {\bibfnamefont {N.}~\bibnamefont {Shiga}}, \bibinfo
  {author} {\bibfnamefont {W.~M.}\ \bibnamefont {Itano}}, \ and\ \bibinfo
  {author} {\bibfnamefont {J.~J.}\ \bibnamefont {Bollinger}},\ }\href@noop {}
  {\bibfield  {journal} {\bibinfo  {journal} {Nature}\ }\textbf {\bibinfo
  {volume} {458}},\ \bibinfo {pages} {996} (\bibinfo {year}
  {2009})}\BibitemShut {NoStop}%
\bibitem [{\citenamefont {Myatt}\ \emph {et~al.}(2000)\citenamefont {Myatt},
  \citenamefont {King}, \citenamefont {Turchette}, \citenamefont {Sackett},
  \citenamefont {Kielpinski}, \citenamefont {Itano}, \citenamefont {Monroe},\
  and\ \citenamefont {Wineland}}]{myatt}%
  \BibitemOpen
  \bibfield  {author} {\bibinfo {author} {\bibfnamefont {C.~J.}\ \bibnamefont
  {Myatt}}, \bibinfo {author} {\bibfnamefont {B.~E.}\ \bibnamefont {King}},
  \bibinfo {author} {\bibfnamefont {Q.~A.}\ \bibnamefont {Turchette}}, \bibinfo
  {author} {\bibfnamefont {C.~A.}\ \bibnamefont {Sackett}}, \bibinfo {author}
  {\bibfnamefont {D.}~\bibnamefont {Kielpinski}}, \bibinfo {author}
  {\bibfnamefont {W.~M.}\ \bibnamefont {Itano}}, \bibinfo {author}
  {\bibfnamefont {C.}~\bibnamefont {Monroe}}, \ and\ \bibinfo {author}
  {\bibfnamefont {D.~J.}\ \bibnamefont {Wineland}},\ }\href@noop {} {\bibfield
  {journal} {\bibinfo  {journal} {Nature}\ }\textbf {\bibinfo {volume} {403}},\
  \bibinfo {pages} {269} (\bibinfo {year} {2000})}\BibitemShut {NoStop}%
\bibitem [{\citenamefont {Teklemariam}\ \emph {et~al.}(2003)\citenamefont
  {Teklemariam}, \citenamefont {Fortunato}, \citenamefont {L{\'o}pez},
  \citenamefont {Emerson}, \citenamefont {Paz}, \citenamefont {Havel},\ and\
  \citenamefont {Cory}}]{cory}%
  \BibitemOpen
  \bibfield  {author} {\bibinfo {author} {\bibfnamefont {G.}~\bibnamefont
  {Teklemariam}}, \bibinfo {author} {\bibfnamefont {E.}~\bibnamefont
  {Fortunato}}, \bibinfo {author} {\bibfnamefont {C.}~\bibnamefont
  {L{\'o}pez}}, \bibinfo {author} {\bibfnamefont {J.}~\bibnamefont {Emerson}},
  \bibinfo {author} {\bibfnamefont {J.~P.}\ \bibnamefont {Paz}}, \bibinfo
  {author} {\bibfnamefont {T.}~\bibnamefont {Havel}}, \ and\ \bibinfo {author}
  {\bibfnamefont {D.}~\bibnamefont {Cory}},\ }\href@noop {} {\bibfield
  {journal} {\bibinfo  {journal} {Phys. Rev. A}\ }\textbf {\bibinfo {volume}
  {67}},\ \bibinfo {pages} {062316} (\bibinfo {year} {2003})}\BibitemShut
  {NoStop}%
\bibitem [{\citenamefont {Biercuk}\ \emph {et~al.}(2011)\citenamefont
  {Biercuk}, \citenamefont {Doherty},\ and\ \citenamefont {Uys}}]{biercuk}%
  \BibitemOpen
  \bibfield  {author} {\bibinfo {author} {\bibfnamefont {M.}~\bibnamefont
  {Biercuk}}, \bibinfo {author} {\bibfnamefont {A.}~\bibnamefont {Doherty}}, \
  and\ \bibinfo {author} {\bibfnamefont {H.}~\bibnamefont {Uys}},\ }\href@noop
  {} {\bibfield  {journal} {\bibinfo  {journal} {J. Phys. B}\ }\textbf
  {\bibinfo {volume} {44}},\ \bibinfo {pages} {154002} (\bibinfo {year}
  {2011})}\BibitemShut {NoStop}%
\bibitem [{\citenamefont {Zurek}(1982)}]{zurek}%
  \BibitemOpen
  \bibfield  {author} {\bibinfo {author} {\bibfnamefont {W.~H.}\ \bibnamefont
  {Zurek}},\ }\href@noop {} {\bibfield  {journal} {\bibinfo  {journal} {Phys.
  Rev. D}\ }\textbf {\bibinfo {volume} {26}},\ \bibinfo {pages} {1862}
  (\bibinfo {year} {1982})}\BibitemShut {NoStop}%
\bibitem [{\citenamefont {Yuge}\ \emph {et~al.}(2011)\citenamefont {Yuge},
  \citenamefont {Sasaki},\ and\ \citenamefont {Hirayama}}]{yuge}%
  \BibitemOpen
  \bibfield  {author} {\bibinfo {author} {\bibfnamefont {T.}~\bibnamefont
  {Yuge}}, \bibinfo {author} {\bibfnamefont {S.}~\bibnamefont {Sasaki}}, \ and\
  \bibinfo {author} {\bibfnamefont {Y.}~\bibnamefont {Hirayama}},\ }\href@noop
  {} {\bibfield  {journal} {\bibinfo  {journal} {Phys. Rev. Lett.}\ }\textbf
  {\bibinfo {volume} {107}},\ \bibinfo {pages} {170504} (\bibinfo {year}
  {2011})}\BibitemShut {NoStop}%
\bibitem [{\citenamefont {{\'A}lvarez}\ and\ \citenamefont
  {Suter}(2011)}]{alvarez}%
  \BibitemOpen
  \bibfield  {author} {\bibinfo {author} {\bibfnamefont {G.~A.}\ \bibnamefont
  {{\'A}lvarez}}\ and\ \bibinfo {author} {\bibfnamefont {D.}~\bibnamefont
  {Suter}},\ }\href@noop {} {\bibfield  {journal} {\bibinfo  {journal} {Phys.
  Rev. Lett.}\ }\textbf {\bibinfo {volume} {107}},\ \bibinfo {pages} {230501}
  (\bibinfo {year} {2011})}\BibitemShut {NoStop}%
\bibitem [{\citenamefont {Ernst}(2004)}]{ernst}%
  \BibitemOpen
  \bibfield  {author} {\bibinfo {author} {\bibfnamefont {R.}~\bibnamefont
  {Ernst}},\ }\href@noop {} {\bibfield  {journal} {\bibinfo  {journal} {J.
  Chem. Phys.}\ }\textbf {\bibinfo {volume} {45}},\ \bibinfo {pages} {3845}
  (\bibinfo {year} {2004})}\BibitemShut {NoStop}%
\bibitem [{\citenamefont {Ajoy}\ \emph {et~al.}(2011)\citenamefont {Ajoy},
  \citenamefont {{\'A}lvarez},\ and\ \citenamefont {Suter}}]{ashok}%
  \BibitemOpen
  \bibfield  {author} {\bibinfo {author} {\bibfnamefont {A.}~\bibnamefont
  {Ajoy}}, \bibinfo {author} {\bibfnamefont {G.~A.}\ \bibnamefont
  {{\'A}lvarez}}, \ and\ \bibinfo {author} {\bibfnamefont {D.}~\bibnamefont
  {Suter}},\ }\href@noop {} {\bibfield  {journal} {\bibinfo  {journal} {Phys.
  Rev. A}\ }\textbf {\bibinfo {volume} {83}},\ \bibinfo {pages} {032303}
  (\bibinfo {year} {2011})}\BibitemShut {NoStop}%
\bibitem [{\citenamefont {Roy}\ \emph {et~al.}(2011)\citenamefont {Roy},
  \citenamefont {Mahesh},\ and\ \citenamefont {Agarwal}}]{soumyaudd}%
  \BibitemOpen
  \bibfield  {author} {\bibinfo {author} {\bibfnamefont {S.~S.}\ \bibnamefont
  {Roy}}, \bibinfo {author} {\bibfnamefont {T.~S.}\ \bibnamefont {Mahesh}}, \
  and\ \bibinfo {author} {\bibfnamefont {G.~S.}\ \bibnamefont {Agarwal}},\
  }\href {\doibase 10.1103/PhysRevA.83.062326} {\bibfield  {journal} {\bibinfo
  {journal} {Phys. Rev. A}\ }\textbf {\bibinfo {volume} {83}},\ \bibinfo
  {pages} {062326} (\bibinfo {year} {2011})}\BibitemShut {NoStop}%
\bibitem [{\citenamefont {Pan}\ \emph {et~al.}(2011)\citenamefont {Pan},
  \citenamefont {Xi},\ and\ \citenamefont {Gong}}]{pan}%
  \BibitemOpen
  \bibfield  {author} {\bibinfo {author} {\bibfnamefont {Y.}~\bibnamefont
  {Pan}}, \bibinfo {author} {\bibfnamefont {Z.-R.}\ \bibnamefont {Xi}}, \ and\
  \bibinfo {author} {\bibfnamefont {J.}~\bibnamefont {Gong}},\ }\href@noop {}
  {\bibfield  {journal} {\bibinfo  {journal} {J. Phys. B}\ }\textbf {\bibinfo
  {volume} {44}},\ \bibinfo {pages} {175501} (\bibinfo {year}
  {2011})}\BibitemShut {NoStop}%
\end{thebibliography}%
\end{document}